\documentclass[prd,aps,twocolumn,amsmath,amssymb,nofootinbib,preprintnumbers]
{revtex4}

\usepackage{graphicx}
\usepackage{dcolumn}
\usepackage{bm}
\usepackage{amsmath}
\usepackage{amsfonts}
\usepackage{euscript,bbm}
\usepackage{ifthen}
\usepackage{psfrag}

\def\ls{\mathrel{\lower4pt\vbox{\lineskip=0pt\baselineskip=0pt
           \hbox{$<$}\hbox{$\sim$}}}}
\def\gs{\mathrel{\lower4pt\vbox{\lineskip=0pt\baselineskip=0pt
           \hbox{$>$}\hbox{$\sim$}}}}
\def\drawbox#1#2{\hrule height#2pt

\hbox{\vrule width#2pt height#1pt \kern#1pt
              \vrule width#2pt}
              \hrule height#2pt}

\def\Asym#1#2{\vcenter{\vbox{\drawbox{#1}{#2}
              \kern-#2pt       
              \drawbox{#1}{#2}}}}

\def\nn{\nonumber}

\newcommand{\be}{\begin{equation}}
\newcommand{\ee}{\end{equation}}
\newcommand{\bea}{\begin{eqnarray}}
\newcommand{\eea}{\end{eqnarray}}

\begin{document}

%
\title{Low-scale Inflation and Supersymmetry Breaking in Racetrack Models}

\author{Rouzbeh Allahverdi$^{1}$}
\author{Bhaskar Dutta$^{2}$}
\author{Kuver Sinha$^{2}$}

\affiliation{$^{1}$~Department of Physics \& Astronomy, University of New Mexico, Albuquerque, NM 87131, USA \\
$^{2}$~Department of Physics, Texas A\&M University, College Station, TX 77843-4242, USA}


\begin{abstract}

In many moduli stabilization schemes in string theory, the scale of inflation appears to be of the same order as the scale of supersymmetry breaking. For low-scale supersymmetry breaking, therefore, the scale of inflation should also be low, unless this correlation is avoided in specific models. We explore such a low-scale inflationary scenario in a racetrack model with a single modulus in type IIB string theory. Inflation occurs near a point of inflection in the K\"ahler modulus potential. Obtaining acceptable cosmological density perturbations leads to the introduction of magnetized $D7$-branes sourcing non-perturbative superpotentials. The gravitino mass, $m_{3/2}$, is chosen to be around 30 TeV, so that gravitinos that are produced in the inflaton decay do not affect big-bang nucleosynthesis. Supersymmetry is communicated to the visible sector by a mixture of anomaly and modulus mediation. We find that the two sources contribute equally to the gaugino masses, while scalar masses are decided mainly by anomaly contribution. This happens as a result of the low scale of inflation and can be probed at the LHC.

\end{abstract}
MIFTP-09-48\\ December, 2009
\maketitle


\section{Introduction}

Inflation is the dominant paradigm of the early universe cosmology to solve the problems of the hot big-bang model and create the seeds for structure formation. Although observations strongly indicate that a period of superluminal expansion happened~\cite{Komatsu:2008hk}, a successful realization of inflation within high energy physics has remained as a challenge. There have been intensive efforts in recent years for realistic embedding of inflation in particle physics and string theory, so that the scalar field responsible for inflation, the inflaton, has a natural place in the observable or a hidden sector. A low-scale model of inflation implemented in a realistic extension of the standard model (SM) of particle physics may in addition make direct connection between cosmology and phenomenology, which can be explored at the LHC. Such models have been studied in the observable sector, most notably inflation in the minimal supersymmetric standard model (MSSM)~\cite{MSSM} and its minimal extensions~\cite{Sneutrino}.

In string theory models of inflation, the inflaton is a modulus field and belongs to a hidden sector. Many models have been studied, with inflaton candidates from the open string sector as well as the closed string sector (for a comprehensive review, see~\cite{McAllister:2007bg}, \cite{Baumann:2009ni}).

Interestingly, in most moduli stabilization schemes that have been studied, the scale of inflation appears to be correlated with the scale of supersymmetry breaking. This was first pointed out in a class of models in which the inflaton is identified with a K\"ahler modulus \cite{Kallosh:2004yh}. The setting is a KKLT-type compactification in type IIB string theory~\cite{Kachru:2003aw}, with the volume modulus as an inflaton. The scale of supersymmetry breaking is determined by the depth of the AdS vacuum; the Hubble scale during inflation $H_{\rm inf}$ is determined by the height of the barrier which protects the dS vacuum after uplift. These two are typically of the same order, and one thus has $H_{\rm inf} \lesssim m_{{3/2}}$.
This kind of correlation appears to be quite robust when string inflation models are embedded in moduli stabilization schemes. Some recent models where such a relation appears include \cite{Silverstein:2008sg}, \cite{McAllister:2008hb}, \cite{Chen:2009nk}, \cite{Conlon:2008cj}.

This leads to the possibility that the physics of moduli stabilization and string cosmology may leave its imprint on particle physics, in particular the details of soft masses in the visible sector. At first sight, such a correlation is discouraging, since in most models inflation occurs at a high scale, while phenomenological considerations usually prefer low-scale supersymmetry breaking. Taking the gravitino mass to be in the $1$ TeV range, this means that inflation happens many orders of magnitude below the usual GUT scale inflationary scenario. On the other hand, high-scale inflation implies a large gravitino mass and correspondingly massive superpartners. In that case, the usual solution to the hierarchy problem through the use of supersymmetry becomes less attractive. There have been several recent efforts to disentangle inflation and supersymmetry breaking, and construct high-scale inflationary models in string theory that incorporate low-scale supersymmetry breaking \cite{Chen:2009nk}, \cite{Conlon:2008cj}, \cite{Badziak:2008yg}.

In this paper we will pursue the line that it is natural to explore low-scale inflation, given the correlation $H_{\rm inf} \lesssim m_{{3/2}}$, if one accepts that supersymmetry is broken at a low scale.

As an example, we will work out a low-scale inflation model in type IIB string theory, on a Calabi-Yau with a single K\"ahler modulus whose real part will be the inflaton. The scale of supersymmetry breaking and inflation will be taken to be
\begin{equation}
H_{\rm inf} \sim m_{3/2} \sim 30 - 50 ~ {\rm TeV}.
\end{equation}
This specific scale avoids the cosmological gravitino problem. It is known that in models where the inflaton is a modulus, its decay typically results in non-thermal overproduction of gravitinos~\cite{Yanagida}. For $m_{3/2} \sim {\cal O}({\rm TeV})$, gravitinos thus produced decay after Big-Bang Nucleosynthesis (BBN), destroying its successful predictions of the primordial abundance of light elements~\cite{Moroi:1999zb}. A standard solution to this problem is to take $m_{3/2} \gs 30$ TeV.

We will assume that the visible sector is sequestered from the SUSY breaking sector. A combination of anomaly mediation and modulus mediation \cite{Randall:1998uk}, \cite{Giudice:1998xp}, \cite{Choi} then gives the low-energy spectrum of the superpartners. Since the scale of supersymmetry breaking is correlated to inflation, we will find that the pattern of soft masses is constrained by cosmological observables, which can be tested at the LHC. Specifically, we will find that while anomaly and modulus mediations contribute equally to gaugino masses, the scalar masses are decided mainly by anomaly contributions.

The setting for our example will be type IIB racetrack models with fluxes, which have superpotentials of the type $W = W_{\rm flux} + A e^{-a {\rm Re}T} + B e^{-b {\rm Re}T}$, where $W_{\rm flux}$ comes from  $G_3$ fluxes and $T$ is the single K\"ahler modulus (models with single gaugino condensation sectors like KKLT may have problems with realizing slow-roll parameters due to the shape of the K\"ahler modulus potential, as outlined in \cite{Badziak:2008gv}, and hence we will not consider them here). These racetrack models have two AdS minima along the direction ${\rm Re} T$ prior to uplifting \cite{Kallosh:2004yh}, \cite{BlancoPillado:2004ns}. By fine-tuning background fluxes appropriately, one of the minima can be flattened to obtain an inflection point \cite{Linde:2007jn}, as shown in Figure 1. If the field starts very close to the inflection point with negligible kinetic energy, one may obtain an acceptable inflationary model.

Usually, racetrack inflation has been studied with $m_{3/2}$ and $H_{\rm inf}$ around $10^{8}$ GeV or higher \cite{Conlon:2005jm}, \cite{BlancoPillado:2009nw}, \cite{BlancoPillado:2006he}, \cite{Brax:2007fe}, \cite{Lalak:2005hr}. Inflation at lower scales $\sim 1$ TeV in such models typically gives an unacceptably low value of density perturbations, related to the difficulty in obtaining sufficiently small values of the slow-roll parameter $\epsilon$ for natural values of supergravity input parameters.

Within the context of our example, this general difficulty can be traced to the fact that input parameters are either fixed by the gravitino mass, or are not fully calculable and generally taken to be $\mathcal{O}(1)$. It will turn out that to obtain the correct amplitude of density perturbations at a low Hubble scale, the third derivative of the potential near the inflection point has to be large, which is difficult given the above constraints.

The main tool we will employ is to turn on magnetic flux on $D7$-branes sourcing the non-perturbative superpotential, which will be useful in tuning $\epsilon$. Magnetized $D$-branes have previously been studied in scenarios of high-scale inflation \cite{Badziak:2008gv}, \cite{Abe:2008xu}, \cite{Abe:2005pi}.

We note that our purpose is not to construct a globally consistent model in an explicit Calabi-Yau compactification. We will also not calculate potentially destabilizing string loop corrections. However, some essential tools in our example, like the use of magnetized branes, may be general features in low-scale inflation models. It would also be interesting to explore the striking effect of cosmological observables on the pattern of soft masses in our example for other models.

The rest of the paper is oraganized as follows. In section 2, we review the setting of our main example, racetrack inflation in type IIB with background and brane flux. We relegate certain details to the appendix. In section 3, we present our results for low-scale inflation in such models. In section 4, we discuss the mediation of supersymmetry breaking and the mass spectrum of superparticles. We close the paper with conclusions in section 5.


\section{The Model: Single K\"ahler Modulus Racetrack with Brane and Background Flux}

In this section, we review the string theoretic setting for the inflationary scenario described in the Introduction. The essential elements in a KKLT-type model of string compactification are: $(1)$ background fluxes on a type IIB Calabi-Yau three-fold giving a Gukov-Vafa-Witten superpotential contribution, and $(2)$ gaugino condensation on $D7$-branes or Euclidean D3 instantons giving a non-perturbative superpotential contribution. These two contributions are sufficient to stabilize complex structure moduli and the dilaton, as well as K\"ahler moduli, in an AdS vacuum. An additional contribution to the scalar potential coming from anti-D3-branes then lifts the solution to a de Sitter vacuum. The superpotential contribution due to 3-form fluxes $G_3$ is of the form
\be
W_{\rm flux}=\int_{\rm CY_3}G_3\wedge \Omega \,\, .
\ee
On the other hand, gaugino condensation of pure Super-Yang-Mills on a stack of $D7$ branes gives a non-perturbative superpotential
\begin{equation}
W_{\rm np} = A e^{-a f_g}\,\, .
\end{equation}
We will be working in units of the reduced Planck mass $M_{\rm P} = 2.4 \times 10^{18}$ GeV for the remainder of the paper unless explicitly stated otherwise. Here, $f_g$ denotes the $D7$ gauge kinetic function, $a = \frac{2\pi}{N_c}$, and $N_c$ is the rank of the gauge group. $A$ is a function of complex structure moduli, the dilaton, and open string fields. $A$ will be taken to be order one, and its precise dependence on moduli comes through one-loop threshold corrections to the $D7$ gauge kinetic function and higher curvature corrections on the world volume of the $D7$s. In the case when magnetic flux on the $D7$ branes is turned off, one has $f_g = {\rm Re }T$, where $T$ is the complex K\"ahler modulus of the 4-cycle wrapped by the $D7$ branes.

We will turn on magnetic flux on the world volume of the $D7$ brane. This leads to a modification of the gauge kinetic function with an extra magnetic flux-dependent dilaton contribution, which will be useful for our purposes. The details of the construction are given in the Appendix.

With magnetic flux the superpotential has the following form
\begin{equation}\label{finalW}
W_{\rm np} = A e^{-a f_g} = A e^{a\mathfrak{f}_\Sigma {\rm Re} S}e^{-a{\rm Re} T }.
\end{equation}
Here, $\mathfrak{f}_\Sigma$ is a magnetic flux-dependent paramter whose form we give in the Appendix, and $\Sigma$ is the four-cycle wrapped by the $D7$-brane. We will assume that the dilaton $S$ has been fixed by background $G_3$ fluxes.

For gaugino condensation with a group $SU(N_{c1}) \times SU(N_{c2})$ we thus obtain an effective supergravity theory with
\bea \label{setup}
K &=& - 3 \ln(T + \overline{T})\;,\qquad T=\sigma+i\tau \ ,  \nonumber \\
W &=& W_{\rm flux} + Ae^{a\mathfrak{f}_\Sigma {\rm Re} S}e^{-a{\rm Re}T}+ Be^{b\mathfrak{f}_\Sigma {\rm Re} S}e^{-b{\rm Re}T}\,\, ,
\eea
with $a = \frac{2\pi}{N_{c1}}$ and $b = \frac{2\pi}{N_{c2}}$.

As in KKLT, we will use anti-D3 branes for generating the uplifting potential to obtain a dS vacuum. The uplifting potential from anti-D3 branes is given by
\begin{equation} \label{lifting}
V_{\overline{D3}} =  \frac{C}{({\rm Re} (T))^2}
\end{equation}
where $C$ is a coefficient determined by the tension of the $\overline{D3}$.


\section{Low-scale Inflation along the K\"ahler Modulus}

Our effective four-dimensional supergravity theory is given by Eq.~(\ref{setup}). The scalar potential is given by
\be \label{parameters}
V= e^{K}\left(  G^{T \overline T} D_{T}W \overline {D_{T} W}
- 3|W|^2 \right) + \frac{C}{({\rm Re} (T))^2}.
\ee
The minimum in the axion direction lies at ${\rm Im}T \,= \,0$,  provided that $A,a,B,b$ and $W_{\rm flux}$ are real, $a>b$ and $A>|B|$, and $AB < 0$. There are two supersymmetric AdS minima along ${\rm Re}T = \sigma$ given approximately by
\be \label{sigma1}
\sigma_1 \,\sim \, \mathfrak{f}_\Sigma {\rm Re} S + \frac{1}{a-b}\ln \left |\frac{a\,A}{b\,B}\right|
\ee
and
\be \label{sigma2}
\sigma_2 \, \sim \, \mathfrak{f}_\Sigma {\rm Re} S -\frac{1}{a}\ln(W_{\rm flux}/A)\;.
\ee
By appropriate choice of background $G_3$ fluxes and uplifting potential, $\sigma_1$ is tuned to an inflection point as shown in Figure 1.
\begin{figure}[ht]
\centering
\includegraphics[width=3.5in]{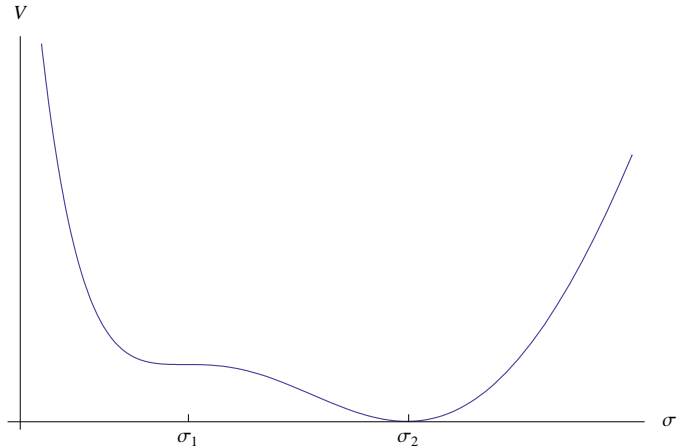}
\caption{The uplifted potential as a function of ${\rm Re}T$ (denoted by $\sigma$). The inflection point at $\sigma_1$ is obtained by tuning background $G_3$ fluxes and the uplifting potential. The global minimum at $\sigma_2$ is a dS vacuum.}
\end{figure}
If $\sigma$ is sufficiently close to $\sigma_1$ and its kinetic energy is negligible, inflation can take place. As mentioned in the introduction, the Hubble scale for such an inflationary scenario should be of the order of the present value of the gravitino mass. The gravitino mass is given by $m^{2}_{3/2} ~=~  \frac{V_{\rm AdS}(\sigma_2)}{3}$ while the Hubble scale is roughly given by $H^2 ~=~  \frac{V_{0}(\sigma_1)}{3}$.

\subsection{Inflection point inflation}

It is useful to expand the potential in the vicinity of the inflection point. Since the second derivative vanishes at $\sigma_1$, we have
\be
V(\sigma) \approx V_0 \left[1-\lambda_1 (\sigma - \sigma_1) -\frac{\lambda_3}{3} (\sigma - \sigma_1)^3\right]
\ee
where $\lambda_p = (\frac{2\sigma_{1}^2}{3})^{p/2}\partial V(\sigma_{1})/V(\sigma_{1})$. The slow-roll parameters are given as usual by
\bea \label{slowroll}
\epsilon&\equiv& {1 \over 2}  \left({V{,\sigma} \over V}\right)^2 = \frac{ 1}{2}\left[\lambda_1 + \lambda_3 (\sigma - \sigma_1)^2 \right]^2 \nn\\ && \\
\eta&\equiv&  \left({V \over V_{,\sigma \sigma}}\right) = -2 \lambda_3 (\sigma - \sigma_1) \quad,\nn
\eea
The number of e-foldings between the time that perturbations at scales of COBE normalization exit the horizon and the end of inflation, denoted by $N_{\rm COBE}$, obeys the following equation~\cite{LL}
\begin{equation} \label{NCOBE1}
N_{\rm COBE} = 68.5 + {1 \over 4} {\rm ln} \left({V_{\rm hor} \over m^4_{\rm pl}}\right) + {1 \over 4} {\rm ln} \left({V_{\rm hor} \over \rho_{\rm end}}\right) + {1 \over 12} {\rm ln} \left({\rho_{\rm reh} \over \rho_{\rm end}}\right) .
\end{equation}
Here $m_{\rm pl} = 1.2 \times 10^{19}$ GeV is the Planck mass, $V_{\rm hor}$ is the inflaton potential when the perturbation exit the horizon, $\rho_{\rm end}$ is the energy density of the universe at the end of inflation, and $\rho_{\rm reh}$ is the energy density at the end of reheating~\footnote{By this we mean the time when the equation of state of the universe changes from that for matter domination, due to inflaton oscillations, to that from radiation domination.}.

Since inflation occurs near a point of inflection where the potential is very flat, the energy density of the universe is practically constant throughout the whole slow roll epoch (this is unlike, for example, models of chaotic inflation). Therefore $V_{\rm hor} \approx \rho_{\rm end} \approx V_0$ and the third term on the right-hand side of Eq.~(\ref{NCOBE1}) is negligible. The inflaton decay is gravitationally suppressed and its rate is given by
\begin{equation} \label{decay}
\Gamma_{\sigma} = {c \over 2 \pi} {m^3_{\sigma} \over M^2_{\rm P}},
\end{equation}
where $c \sim 1$, $m_{\sigma}$ is the inflaton mass (i.e. modulus mass at the minimum of its potential). This implies that $\rho_{\rm reh} = 3 \Gamma^2_{\sigma} M^2_{\rm P}$. On the other hand, $V_0 = 3 H^2_{\rm inf} M^2_{\rm P}$, where $H_{\rm inf}$ is the Hubble expansion rate during inflation.

As explained earlier, we need $H_{\rm inf} \sim m_{3/2} \sim 30$ TeV in order to have a low scale of supersymmetry breaking while avoiding the (non-thermal) gravitino problem. This results in $V_0 \sim 10^{-28} M^4_{\rm P}$ and $m_{\sigma} \sim 3000$ TeV. Therefore Eq.~(\ref{NCOBE1}) yields
\be \label{NCOBE}
N_{\rm COBE} \simeq 43.
\ee
On the other hand, we have
\be \label{NCMB}
N_{{\rm COBE}} = \int_{\sigma_{\rm COBE}}^{\sigma_{\rm end}} \left[\lambda_1 + \lambda_3 (\sigma - \sigma_1)^2 \right]^{-1} d\sigma \,\, ,
\ee
where $\sigma_{\rm COBE}$ is the field value at which the observationally relevant perturbations exit the horizon, and $\sigma_{\rm end}$ is the field value at which the slow roll conditions are violated and inflation ends. Also, the total number of e-foldings of inflation $N_{\rm tot}$ is given by
\be \label{Ntotal}
N_{{\rm tot}} = \int_{\sigma_1}^{\sigma_{\rm end}} \left[\lambda_1 + \lambda_3 (\sigma- \sigma_1)^2 \right]^{-1} d\sigma \,\, .
\ee
Eqs.~(\ref{NCMB}) and~(\ref{Ntotal}) can be used to eliminate $\sigma_{{\rm COBE}}$ and $\lambda_1$ in favor of $N_{{\rm COBE}}$ and $N_{{\rm tot}}$. The slow-roll parameters can then be expressed as follows
\begin{eqnarray} \label{final}
&& \eta_{\rm COBE} = \frac{\pi}{N_{{\rm tot}}} \cot \left(\frac{\pi N_{{\rm COBE}}}{2 N_{{\rm tot}}}\right) \, \nonumber \\
&& \epsilon_{\rm COBE} = {1 \over 2} \left({\pi \over 2} \right)^4 \left[1 + {\rm cot}^2\left({\pi N_{\rm COBE} \over 2 N_{\rm tot}}\right) \right] \lambda_3^{-2}N_{{\rm tot}}^{-4}\, . \nonumber \\
\,
\end{eqnarray}
The power spectrum of density perturbations is given by~\footnote{Gravitational waves produced during inflation are negligible and cannot be observed in future experiments because of the low scale of inflation.}
\be \label{density}
\Delta_{\cal R}^2 = \frac{1}{4\pi^2}\left(\frac{H^2}{\dot\sigma}\right)_{\sigma_{\rm COBE}}^2 = \frac{V_0}{12\pi^2}\lambda_3^{2}N_{{\rm tot}}^{4},
\ee
where we have used Eq.~(\ref{final}). Note that the low scale of inflation requires that $\epsilon$ be extremely small. The spectral index is given by
\be \label{spectral}
n_s = 1 + 2 \eta - 6 \epsilon = 1-\frac{2\pi}{N_{\rm tot}}\cot\left(\frac{\pi N_{\rm COBE}}{2N_{\rm tot}}\right) \,\,.
\ee

\subsection{Model parameters and numerical results}

At this stage, plugging in the values of observed quantities constrains the parameters $\lambda_1$ and $\lambda_3$. Since $N_{\rm COBE} = 43$, see Eq.~(\ref{NCOBE}), we find from Eq.~(\ref{spectral}) that $N_{\rm tot} \simeq 46-72$ in order for $n_s$ to be within the $2 \sigma$ range $n_s = 0.960^{+0.014+0.029}_{-0.015-0.027}$ allowed by the 5-year WMAP data~\cite{Komatsu:2008hk}. Then, recall that $V_0 \sim 10^{-28}$ (in units of $M^4_{\rm P}$), it turns out from Eq.~(\ref{density}) that obtaining the correct value of $\Delta_{\cal R}^2 \simeq 2.0 \times 10^{-9}$ results in
\be \label{lambda3}
\lambda_3 \sim 10^{7}.
\ee
Obtaining acceptable values of $n_s$, see Eq.~(\ref{Ntotal}), results in:
\be \label{lambda1}
4.76 \times 10^{-11} \leq \lambda_1 \leq 1.66 \times 10^{-10},
\ee
where we have used the fact that $46 \leq N_{\rm tot} \leq 72$ in order for $n_s$ to be in the allowed range.

For $\lambda_3 \sim 10^7$, Eq.~(\ref{slowroll}) leads to $(\sigma_{\rm end} - \sigma_1) \sim 10^{-7} \sigma_1$ (note that $\vert \eta \vert \sim 1$ at the end of inflaiton). This implies that the model belongs to the class of small-field models~\cite{McAllister:2007bg},~\cite{Baumann:2009ni} and does not suffer from the problems of Planckian displacements.

One comment is in order at this point. Above, we have assumed that inflation is always in the slow roll regime all the way between $\sigma_1$ and $\sigma_{\rm end}$. Quantum diffusion becomes important where $V_{,\sigma} < 3 H^2_{\rm inf}/2 \pi$. Therefore it takes over classical slow roll if $\partial V(\sigma_1)$ is very small, i.e., when $\sigma_1$ becomes a saddle point. This would result in a self-reproduction regime around $\sigma_1$, and hence slow roll inflation would start not at $\sigma_1$ but slightly away from it. However for values of $\lambda_1$ that yield acceptable spectral index, see Eq.~(\ref{lambda1}), it turns out that the quantum diffusion is always subdominant and slow roll regime indeed begins at the inflection point $\sigma_1$.

The values of $V_0$, $\lambda_3$, $\lambda_1$ determined from the scale of inflation, amplitude of perturbations and the scalar spectral index, respectively, are translated into constraints on the parameters of the potential~(\ref{parameters}). For a specific example, we use the following input parameters
\bea \label{input}
&& W_{{\rm flux}} = 2 \cdot 10^{-10}, \, A = 1, \, B = -0.0336008908401,\nonumber \\
&& a = \frac{2\pi}{6},\, b=\frac{2\pi}{7}, \,C=1.53\cdot 10^{-22} \nonumber \\
&& a\mathfrak{f}_\Sigma {\rm Re} S = 300
\eea
For this choice, we obtain
\be
\sigma_1 \, = \, 324.54, \;\; \sigma_2 \, = \, 325.25
\ee
We solved the full equation of motion of the inflaton
\be \label{sigmaeom}
\sigma^{\prime \prime} = -\left(1-\frac{\sigma'^2}{4\sigma^2}\right)\left(3\sigma'+2\sigma^2\frac{V_{,\sigma}}{V}\right)+\frac{\sigma'^2}{\sigma}\,\, ,
\ee
where derivatives are with respect to $N = N_{\rm tot} - N_{\rm COBE}$, i.e. the number of e-foldings of inflation from the inflection point $\sigma_1$~\footnote{We note that $\sigma$ does not have a canonical kinetic term. The metric for $\sigma$ is given by $g_{\sigma \sigma} = 3/2 \sigma^2$ as can be seen from the expression for the K\"ahler potential. Thus, all derivatives with respect to $\sigma$ have to be redefined by $\partial_{\sigma} \rightarrow \sqrt{g^{\sigma \sigma}}\partial_{\sigma}$.}. The initial conditions are given as $\sigma(0) = \sigma_1, \, \sigma'(0) = 0$.

The evolution of $\sigma$ as a function of $N$ is plotted in Figure 2. The total number of e-foldings is $N_{\rm tot} = 66$, corresponding to $\eta \sim 1$. The scalar spectral index $n_s$ is plotted as a function of $N$ in Figure 3. At $N = 23$, corresponding to $N_{\rm COBE} = 43$, we have $n_s = 0.97$.

\begin{figure}[ht]
\centering
\includegraphics[width=3.5in]{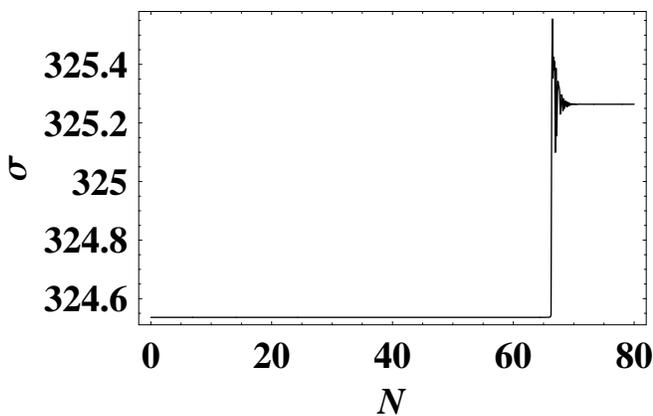}
\caption{The evolution of ${\rm Re}T$ (denoted by $\sigma$ in the figure) as a function of the number of e-foldings $N$. The field inflates for about 66 e-foldings at the inflection point $\sigma_1 = 324.54$. It then starts rolling into the global minimum at $\sigma_2 = 325.25$, first oscillating before stabilizing at the bottom.}
\end{figure}
\begin{figure}[ht]
\centering
\includegraphics[width=3.5in]{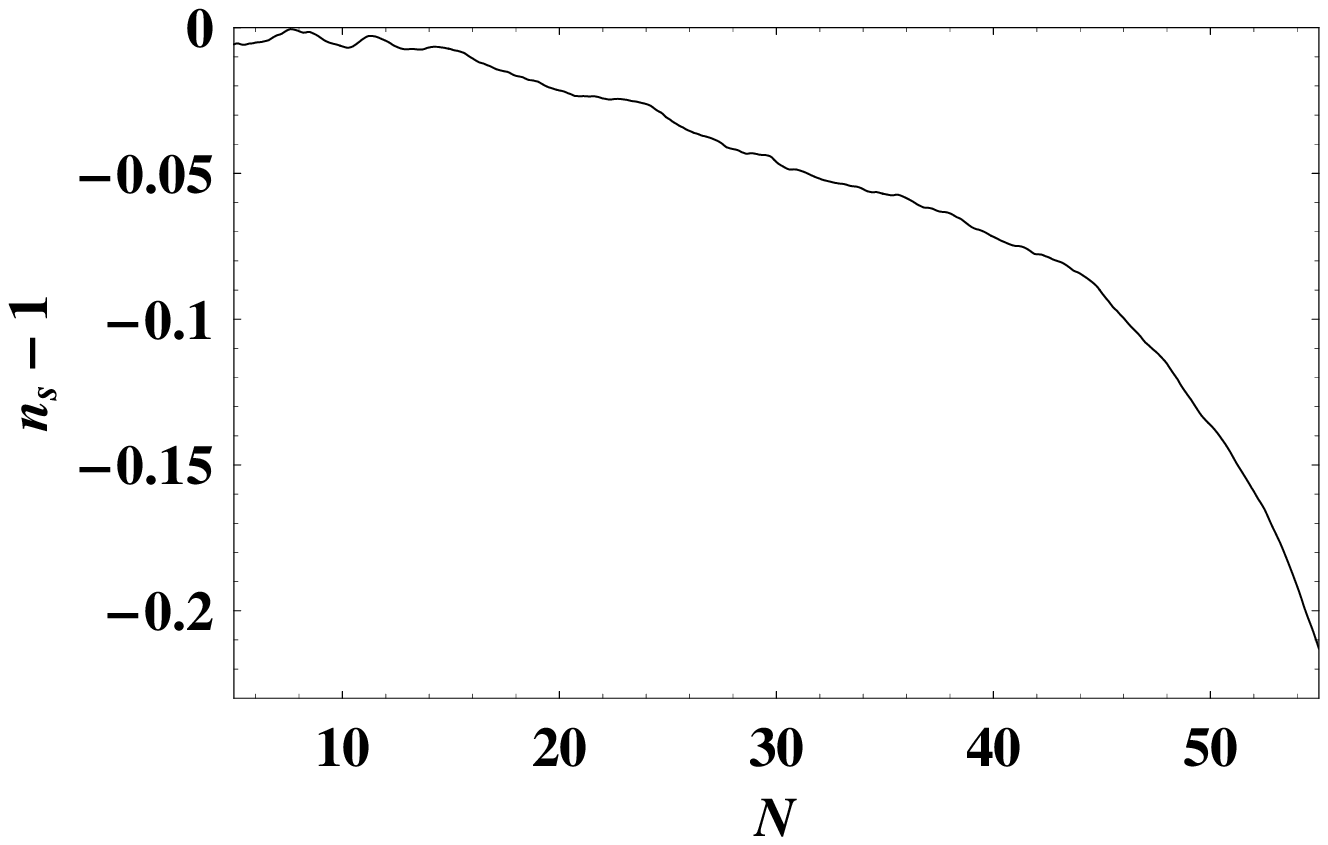}
\caption{The spectral index as a function of the number of e-foldings $N$. $\eta \sim \frac{1}{2}(n_s - 1)$ becomes order $1$ at the end of inflation $N_{\rm tot} = 66$. At $N=23$, the modes at the COBE normalization scale exit the horizon, at which point $n_s = 0.97$.}
\end{figure}
Note that in this class of models, one uplifting potential is used to obtain both an inflection point and a dS vacuum. For the parameters chosen above, the value of V at the global minimum is $V(\sigma_2) \sim 10^{-28}M_p^4$. This value can be lowered in different ways, e.g., by rescaling the existing  parameters, by using an uplifting potential with more parameters or by simply assuming another sector that is not responsible for inflation or SUSY breaking, in a generic compactification with multiple moduli. None of these methods would  affect the discussion of the soft masses, which are fixed by the gravitino mass and relations (\ref{lambda3}), (\ref{lambda3analytical}). It would also be interesting to extend our methods to the multi-moduli case. In this paper, however, we will not be addressing the tuning required to obtain a lower value of V, which we leave for future work.

\subsection{Tuning of parameters}

It is seen from Eq.~(\ref{lambda1}) that there is a tuning of ${\cal O}(10^{-10})$ in $\lambda_1$ to match the allowed range of $n_s$ in our model. This tuning is achieved by adjusting the value of the coefficient $B$ with respect to the coefficient $A = 1$ at the same level. However, the real issue is stability of the fine tuning against higher border stringy corrections~\footnote{Corrections from the observable sector are totally negligible because the inflaton has Planck suppressed couplings to the fields in the observable sector.}.

The K\"ahler potential receives $\alpha^{\prime}$ and string loop corrections. The $\alpha^{\prime}$ corrections \cite{Becker:2002nn} go as
\be
\delta \, K_{\alpha^{\prime}} \, = \, - \frac{\xi}{(T + \overline{T})^{3/2}} + \mathcal{O}(1/(T + \overline{T})^{9/4}) ,
\ee
where the leading order correction comes at $\mathcal{O}(\alpha^{\prime 3})$. Here, $\xi$ is a parameter that depends on the Euler number of the Calabi-Yau.

The corrections to the scalar potential go as inverse powers of the volume of the Calabi-Yau with the first term coming at
\be
\delta \, V_{\alpha^{\prime}}^{(1)} \, \sim \,  \frac{1}{(T + \overline{T})^{27/8}}
\ee
For our value ${\rm Re} T \sim 324$ this is of order $10^{-10}$. Thus, the value of $\lambda_1$ needs to be tuned to first order in $\alpha^{\prime}$ corrections. We neglect higher orders.

String loop corrections to the K\"ahler potential have been discussed for example in \cite{Cicoli:2009zh} and derived in detail in the case of toroidal orientifolds without fluxes \cite{Berg:2007wt}. For arbitrary Calabi-Yau compactifications, the leading contribution of loop corrections to the scalar potential vanishes, giving an extended no-scale structure, only if the corrections satisfy certain conditions \cite{Cicoli:2007xp}. The first non-vanishing corrections to the scalar potential in such cases are subdominant to the $\alpha^{\prime}$ correction. In general Calabi-Yau compactifications, string loop corrections may reintroduce the $\eta$ problem. We will not consider them any further in this paper.

Let us compare the amount of fine tuning in our model with that of low-scale inflection point inflation in the observable sector~\cite{MSSM}. The tree-level fine tuning shows an improvement by several order of magnitude in this case, which comes as a direct consequence of the larger scale of inflation, $H_{\rm inf} \sim 30$ TeV in the former compared with $H_{\rm inf} \sim 100$ MeV in the latter. The stability of the fine tuning is also better in this case since higher order corrections to the potential parameters are much smaller, ${\cal O}(10^{-10})$ in the former vs ${\cal O}(10^{-2})$ in the latter where the leading order corrections come from one-loop diagrams with gauge strength interactions.

\subsection{Role of magnetic flux}

The most important constraint on supergravity parameters comes from Eq.~(\ref{lambda3}). This can be seen by explicitly working out the $\lambda_p$ from the scalar potential Eq.~(\ref{parameters}).

It is convenient to simplify the scalar potential using $a \sim b$ and $A = 1$, obtaining
\bea
V &\sim& \frac{1}{6\sigma^2} e^{-2a\sigma} \times \nonumber \\
&& \left(6C e^{2a\sigma} +
   a(1+B)^2 e^{2 a\mathfrak{f}_\Sigma {\rm Re} S}(3 + a\sigma)\right.\nonumber \\
&& \left.+
   3a(1+B)e^{a\mathfrak{f}_\Sigma {\rm Re} S + a\sigma}W_{{\rm flux}} \right)
\eea
We note that the uplifting coefficient $C$ and $W_{{\rm flux}}$ may be roughly fixed by the inflection point condition and the gravitino mass. It is easy to check that $\lambda_1 \sim 10^{-10}$ translates roughly to a quadratic equation in $B$, with coefficients that are functions of $a\sigma$. For a given $a\sigma$, one can tune $B$ appropriately to obtain the required $\lambda_1$, and it turns out that $B$ is also $\mathcal{O}(1)$ and negative.

On the other hand, for $\lambda_3$ one obtains the leading order behavior
\be \label{lambda3analytical}
\lambda_3 \sim a^3\sigma^3 \,\,.
\ee
From Eq.~(\ref{sigma2}), we thus obtain $\lambda_3 \sim  (a \mathfrak{f}_\Sigma {\rm Re} S - \ln W)^3$.

This is a strong constraint. In the absence of magnetic brane flux $\mathfrak{f}_\Sigma$, the natural value is $a\sigma \sim 10-20$, a result that only depends logarithmically on the scale of supersymmetry breaking. To match the magnitude of primordial curvature perturbations from Eq.~(\ref{density}), then, one obtains high-scale inflation with $H_{\rm inf} \sim 10^9$ GeV. Thus, non-magnetic models naturally have a high scale of inflation and supersymmetry breaking.

We thus see that to obtain the correct value of $\lambda_3 \sim 10^7$ and match the observed value of curvature perturbations, we require a non-trivial magnetic flux dependent term
\be \label{fS}
a \mathfrak{f}_\Sigma {\rm Re} S \,\, \sim \,\, 300 \,.
\ee
The expression for $\mathfrak{f}_\Sigma {\rm Re} S$ in terms of magnetic flux quanta and the underlying geometry is discussed in the Appendix. It may be possible to obtain (\ref{fS}), however global consistency conditions such as the cancellation of RR tadpoles must be checked in an explicit construction. This is beyond the scope of this paper.


\section{Supersymmetry Breaking and the Superparticle Spectrum}

In this section, we discuss the supersymmetry breaking pattern in this model of low-scale inflation.

The visible sector is assumed to be sequestered from the supersymmetry breaking brane \cite{Kachru:2007xp}. The soft masses are determined by a combination of modulus mediation that is $\mathcal{O}(F^T/T)$, and anomaly mediation that is $\mathcal{O}(m_{3/2}/16\pi^2)$. For details of mirage mediation,  the reader may refer to \cite{Choi}. The model is given by
\bea
K &=& - 3 \ln(T + \overline{T})+Z_i(T+T^*)\Phi^*_i\Phi_i, \nonumber \\
W &=& W_{\rm flux} + A e^{-a f_g}+ B e^{-b f_g} +\frac{1}{6}\lambda_{ijk} \Phi_i\Phi_j\Phi_k,
\nonumber
\\
f_g &=&  {\rm Re} T -
\mathfrak{f}_\Sigma {\rm Re} S\,  \eea
where $\Phi_i$ are the visible sector superfields and $Z_i(T+T^*) = 1/(T+T^*)^{n_i}$, $n_i$ being the modular weight.

The soft parameters at the GUT scale are given by
\bea \label{softGUT}
M_a & = & M_0 \Big[\,1+\frac{\ln(M_{\rm P}/m_{3/2})}{16\pi^2} b_ag_a^2\alpha\,\Big],
\nonumber \\
\mathcal{A}_{ijk} & = & M_0\Big[\,(\mathcal{A}_i+\mathcal{A}_j+\mathcal{A}_k) \hspace{2cm}
\nonumber \\
& - & \frac{\ln(M_{\rm P}/m_{3/2})}{16\pi^2}(\gamma_i+\gamma_j+\gamma_k)\alpha\,\Big],
\nonumber \\
m_i^2 & = & M_0^2\Big[\,c_i-\,\frac{\ln(M_{\rm P}/m_{3/2})}{16\pi^2}
\theta_i\alpha \hspace{1cm}
\nonumber \\
& - & \left(\frac{\ln(M_{Pl}/m_{3/2})}{16\pi^2}\right)^2\dot{\gamma}_i\alpha^2\,\Big] \, .
\eea
Here, $b_a$ are the one-loop beta functions, $\gamma_i$ and $\dot{\gamma}_i$ the anomalous dimensions and their derivatives, and $\theta_i$ functions of the quadratic Casimirs and normalized Yukawas of the visible sector.

The parameters $\alpha$ (which measures the ratio of anomaly to modulus contributions), $\mathcal{A}_i,$ and $c_i$ are defined as follows.
\bea
\alpha\,\equiv\,\frac{m_{3/2}}{M_0\ln(M_{\rm P}/m_{3/2})},\quad
\mathcal{A}_i\,\equiv\,\frac{\tilde{\mathcal{A}}_{i}}{M_0}, \quad
c_i\,\equiv\, \frac{\tilde{m}_i^2}{M_0^2}.
\label{eq:def}
 \eea
where $M_0, \tilde{\mathcal{A}}_{ijk},$ and $\tilde{m}_i$ are pure modulus contributions to gaugino masses, trilinear couplings, and sfermion masses, given as functions of the modulus $T$. These modulus contributions are given by the following expressions.
\bea
\label{tmediation} M_0 &=& F^T\partial_T\ln{\rm Re}(f_g),
\nonumber \\
\tilde{m}_i^2 &=& -F^TF^{T*}\partial_T\partial_{\bar{T}}
\ln(e^{-K_0/3}Z_i),
\nonumber \\
\tilde{\mathcal{A}}_{ijk} &=& F^T\partial_T\ln(e^{-K_0}Z_iZ_jZ_k)
\nonumber \\
\eea
where $K_0$ is given by Eq.~(\ref{setup}).

The input parameters for RG running are thus $\alpha,\,\, \mathcal{A},\,\, c,\,\, $tan$\beta$,\,\, and $M_0$ (or equivalently $m_{3/2}$).

It is instructive to compute the above parameters in terms of our underlying string construction. We find
\be \label{alphafull}
\alpha = - \frac{{\rm Re}(f_g)}{\ln(M_{\rm P}/m_{3/2})} \frac{W_{TT}}{W_T}
\ee
where
\be
\frac{W_{TT}}{W_T} = \frac{3 a^2 W_{\rm {flux}} + B e^{-b f_g}(b-a)ab(2{\rm Re} T)}{3a W_{\rm {flux}} + 3B e^{-b f_g}(b-a)}.
\ee
In different limits, either anomaly or modulus contribution will dominate.

$(1).$ $W_{flux} = 0$ and magnetic flux $\mathfrak{f}_\Sigma = 0$. The scenario reduces to pure racetrack, with $\frac{W_{TT}}{W_T} = ab(T+\overline{T})$ and $\alpha \sim (aT)^2/(\ln(M_{\rm P}/m_{3/2}))$. Anomaly contributions dominate.

$(2).$ $W_{flux} \neq 0$ and $\mathfrak{f}_\Sigma = 0$. In this case, $\frac{W_{TT}}{W_T} \sim a $. Thus, $\alpha \sim \frac{aT}{\ln(M_{\rm P}/m_{3/2})}$ and is typically $\mathcal{O}(1)$ in models of mirage mediation based on KKLT. Anomaly and modulus contributions are roughly similar in such cases.

$(3).$ $W_{flux} \neq 0$ and $\mathfrak{f}_\Sigma \neq 0$. Here too, $\frac{W_{TT}}{W_T} \sim a $ and one has
\be
\alpha \sim \frac{a{\rm Re}(f_g)}{\ln(M_{\rm P}/m_{3/2})} \sim \frac{a{\rm Re}T}{32}\left( 1- \frac{\mathfrak{f}_\Sigma {\rm Re} S}{{\rm Re} T} \right) \,\, .
\ee
This is our scenario and from  Eq.~(\ref{sigma2}), we see that in general $\alpha$ is $\mathcal{O}(1)$. For the parameters considered in the example in this paper, we have $\alpha =  0.8$.

The values of $\mathcal{A}_i$ and $c_i$ are also dependent on the magnetic flux. The exact dependence is
\be
\mathcal{A}_i = (1-n_i)\left(1- \frac{\mathfrak{f}_\Sigma {\rm Re} S}{{\rm Re} T}\right), \,\, c_i = (1-n_i)\left(1- \frac{\mathfrak{f}_\Sigma {\rm Re} S}{{\rm Re} T}\right)^2
\ee
Thus, for modular weights $n_i = 0$,
\be \label{inputfinal}
\alpha \, \sim \frac{a{\rm Re}T}{32}\cdot \mathcal{A} = \frac{a{\rm Re}T}{32}\cdot c^{1/2} \,\,\,\, .
\ee
%
This implies that obtaining acceptable density perturbations leaves its imprint on the low-energy pattern of soft masses. 
Since $a{\rm Re}T \sim 300$ in our model, we obtain the result~\footnote{Note that for high-scale inflation models in the KKLT set up we have $\alpha \sim \mathcal{A} \sim c^{1/2}$.}
\be
\alpha \sim 10 \mathcal{A} \sim 10 c^{1/2} \,\,\, .
\ee
For $\alpha \sim 1$, the gaugino masses receive equal modulus and anomaly contributions. For this example, the gluino mass is about 1.5 TeV and the lightest neutralino mass is about 740 GeV. If we use $\alpha \sim 1.5$, the gluino mass becomes 700 GeV and the lightest neutralino mass becomes 340 GeV. However, the values of $\mathcal{A}$ and $c$ are small and modulus contributions are suppressed for the scalars. In particular, the spectrum has tachyonic sleptons. Various model-building techniques exist in the literature to lift tachyonic directions \cite{Anomaly}.

We leave a detailed study of superparticle mass spectrum for an upcoming publication.


\section{Conclusion}

In this paper, we have presented a simple model of closed string inflation in string theory in which the Hubble scale and the scale of supersymmetry breaking are both low. This has been motivated by the fact that in moduli stabilization schemes, $H_{\rm inf} \sim m_{3/2}$ holds generally. Low-energy supersymmetry, which will be very soon investigated at the LHC, is desirable to solve the hierarchy problem.

We worked out an example of single-field low-scale inflation, $H_{\rm inf} \sim 30-50$ TeV, in Calabi-Yau manifolds with a single K\"ahler modulus. Inflation occurs near a point of inflection in the K\"ahler modulus potential. The moduli stabilization schemes with gaugino condensation on magnetized branes allow a successful implementation of inflection point inflation. The magnetic flux quanta, suitably tuned, can produce the correct amplitude of cosmological density perturbations. The scalar spectral index can take have any value within the whole range allowed by the 5-year WMAP data by tuning the value of input background flux parameters.

The scale of inflation ensures that the gravitino problem will be avoided in this model. Inflaton decay leads to copious non-thermal production of gravitinos that would destroy BBN predictions if we had $m_{3/2} \sim {\cal O}({\rm TeV})$. However, for $m_{3/2} \sim 30-50$ TeV, gravitinos decay before BBN.
The soft masses in this model have both moduli and anomaly mediation contributions. In particular, obtaining acceptable density perturbations implies that the gaugino masses receive comparable modulus and anomaly contributions, whereas the scalar masses mainly receive anomaly contributions. Therefore inflation, although happening in the hidden sector, can have an impact on the observable sector through the distinctive mass spectra that can be investigated at the LHC.

The techniques developed in this paper may be applicable in broader settings, such as models of brane inflation embedded in moduli stabilization schemes \cite{Kachru:2003sx}, models with many K\"ahler moduli, etc. In particular, it would be interesting to work out cosmological imprints on SUSY breaking in such models.

\section{Acknowledgements}

The work of B.D. is supported in part by the DOE grant DE-FG02-95ER40917. The work of K.S. is supported by NSF under grant PHY-0505757, PHY05-51164. We would like to thank Daniel Robbins, Waldemar Schulgin, and Gonzalo Torroba for discussions.


\appendix

\section{Gauge kinetic function on magnetized $D7$ branes}

Here, we discuss the gauge kinetic function on magnetized $D7$ branes, along the lines of \cite{braneflux}. Magnetized D-branes have been widely studied for the construction of semi-realistic string vacua, in the context of type IIB toroidal models and Calabi-Yau compactifications \cite{Diaconescu:2006nk}, \cite{IIBtoroidal}, and, by T-duality, type IIA intersecting brane models \cite{IIAintersecting}.

We consider a Calabi-Yau 3-fold $Y$ with a holomorphic involution $\sigma$ under which the K\"ahler form $J$ is even and the (3,0) form $\Omega$ is odd. There are thus O3 and O7 planes at fixed loci of $\sigma$, after modding out by the usual orintifold action. There is a $D7$ brane on a 4-cycle $\Sigma$.

On the $D7$ brane, we turn on a $U(1)$ magnetic flux, on a 2-cycle which is in the 2-homology of $\Sigma$. We will assume that the $D7$ brane does not intersect any stacks of D3 branes, and that the Calabi Yau, the holomorphic involution, magnetic flux, and brane $4$-cycle can be chosen to cancel RR tadpoles. The DBI action for the D-brane is given in the string frame by
\begin{equation} \label{DBI}
S_{DBI}=-\mu_7\int d^{8}\xi e^{-\phi}\sqrt{-\det(\iota^*g
+\iota^*B+2\pi\alpha' F)}\ ,
\end{equation}
where the integral is over the eight dimensional world-volume, $\iota^*g$ and $\iota^*B$  are the pullbacks of the ten-dimensional metric and the NSNS 2-form to the D-brane world-volume, $\mu_7$ is the D-brane tension, $\phi$ is the ten-dimensional dilaton, and $F$ is the field-strength of the $U(1)$ gauge field on the $D7$. Defining $\mathcal F = (\iota^*B+2\pi\alpha' F)|_{\Sigma}$ and using the fact that the divisor $\Sigma$ is holomorphically embedded in the Calabi-Yau, one obtains
\begin{equation}\label{twozero}
\mathcal F^{2,0}=\mathcal F^{0,2}=0\ .
\end{equation}
For simplicity, we take the negative $\sigma$ eigenspaces of the Calabi-Yau to vanish, so that the NSNS 2-form $B$ vanishes. Then, $\mathcal F = 2\pi\alpha' F$, and using the quantization of $F$, one obtains
\begin{equation}\label{quantization}
\int \mathcal F = 4 \pi^2 \alpha' n\ , \quad n \in \mathbb{Z}\ .
\end{equation}

The 2-form flux can be expanded on a basis of harmonic forms on $H^{(1,1)}(\Sigma)$
\begin{equation}\label{DBRANEeq6}
\mathcal F=f^\alpha\iota^*\omega_\alpha+\tilde f^a\tilde\omega_a\ .
\end{equation}
where $\iota^*\omega_\alpha$ are pullbacks of a basis $\omega_\alpha$ of $H^{(1,1)}(Y,\mathbb{Z})$ and $\tilde\omega_a$ are $(1,1)$ harmonic forms on $\Sigma$ that lie on the cokernel of $\iota^*$.

The low energy expansion of the DBI action (\ref{DBI}) gives the gauge kinetic function $f_g$ and a D-term contribution to the scalar potential. It turns out that
\begin{equation}
f_g \,\, \propto \,\, \int_{\Sigma}(\iota^*J\wedge\iota^*J
-\mathcal F\wedge\mathcal F) \ ,\\
\end{equation}
\begin{equation} \label{Dtermflux}
D \,\, \propto \,\,
\int_{\Sigma}\iota^*J\wedge\mathcal F\ .
\end{equation}
It is useful to define a quantity
\begin{equation} \label{fluxparameter}
\mathfrak{f}_\Sigma=\frac12 \Big(f^\alpha f^\beta\mathcal K_{\alpha\beta\Sigma}+\tilde f^a\tilde f^b\mathcal
K_{ab}\Big)
\end{equation}
where $K_{\alpha\beta\Sigma}$ is the triple intersection number of $\Sigma$ and the dual 4-cycles of $\omega_\alpha$ and $\omega_\beta$, while $\mathcal K_{ab}=\alpha^{\prime -2}\int_{\Sigma}\tilde\omega_a\wedge\tilde\omega_b$.

In terms of $\mathfrak{f}_\Sigma$ the gauge kinetic function works out to be
\begin{equation} \label{gaugecoupling}
f_g =  {\rm Re} T -
\mathfrak{f}_\Sigma {\rm Re} S\ .
\end{equation}
Here, $T$ is the K\"ahler modulus of $\Sigma$ and $e^{-\phi}$ is the real part of the dilaton $S$, up to normalization.

In the above analysis, there are also open string moduli corresponding to $D7$ fluctuations and Wilson lines. We assume that they have been fixed by fluxes.

Several comments are in order. First, $\mathfrak{f}_\Sigma$ receives curvature corrections that depend on the first Pontryagin classes of the tangent and normal bundles of $\Sigma$. These corrections in turn feed into the part of the gauge kinetic function that depends on the dilaton. We treat $\mathfrak{f}_\Sigma {\rm Re} S$ as an input parameter that can be varied by changing the magnetic flux number $n$ and suitably choosing $K_{\alpha\beta\Sigma}$ and $\mathcal K_{ab}$. Therefore curvature corrections have been assumed to contribute to its final value.

Second, we mainly deal with Calabi-Yaus with $h^{1,1}=1$, i.e. a single K\"ahler modulus, which is the volume modulus $T$.

Finally, much of the literature on magnetized $D7$ branes in KKLT-type models has focussed on the D-term contribution to the scalar potential, which lifts the AdS vacuum. The D-term potential from magnetic fluxes on $D7$ goes as $\frac{1}{[{\rm Re} (T)]^3}$. This has been explored as an alternative to the introduction of anti-D3 branes to break supersymmetry \cite{braneflux}. The advantage of the magnetized $D7$ brane is that it can be incorporated in a standard supergravity framework, and thus there is control at all stages.

In this work, however, we are interested in the modification of the gauge kinetic function due to magnetic fluxes and its cosmological consequences. We simply use the standard anti-D3 brane picture for uplift, taking magnetic fluxes such that the D-term vanishes in (\ref{Dtermflux}). While the $D$-term contributions can also easily be used to obtain the inflection point, we note that such a procedure will imply more conditions on brane magnetic flux, apart from (\ref{fS}). It would be interesting to investigate this issue further.


\end{document}